# RE-DIFFERENTIATION AS COLLECTIVE INTELLIGENCE: THE KTUNAXA LANGUAGE ONLINE COMMUNITY.

Christopher Horsethief

Doctoral Program in Leadership Studies
Gonzaga University
502 E. Boone Ave, Spokane, WA 99258
e-mail: chorsethief@zagmail.gonzaga.edu

## ABSTRACT

This paper presents preliminary results of an investigation of collectively intelligent behavior in a Native North American speech community. The research reveals several independently initiated strategies organized around the collective problem of language endangerment. Specifically, speakers are engaging in self-organizing efforts to reverse historical language simplification that resulted from cultural trauma. These acts of collective intelligence serve to reduce entropy in speech community identity.

## INTRODUCTION

Bloch and Trager (1942) asserted language was more than an arbitrary system of exchanging symbols. They referred to language as a catalyst for all social activity. It was not merely a component of culture, rather "it is the basis for all cultural activities" (p. 5). This relationship inextricably links language to identity. Consequently, identifying factors impacting language equates to identifying challenges to maintaining community identity. The United Nations Educational, Scientific and Cultural Organization (UNESCO) explored these obstacles in terms of endangerment: a language is endangered when it ceases to be spoken, when speakers utilize fewer registers or variations, and when older generations cease passing it on to future generations. The UNESCO (2010) research classified 43% of the 600 currently spoken languages as endangered. 422 of these are in North America: 191 in the United States, 88 in Canada, and 143 in Mexico.

Crawford (1995) placed the discussion of language loss in the arena of social justice, connecting the loss of language with negative impact on self–worth, poverty, failure of family systems, and – most importantly to the collective – limited potential and problem-solving endeavors. He noted "after all, language death does not happen in privileged communities. It happens to the dispossessed and disempowered, peoples who most need their cultural resources to survive" (pp. 34-35). If we believe that when a language dies its traditional knowledge also dies, and we take the UNESCO statistics to be accurate, then we are faced with the impending loss of 422 cultures in North America. One of these cultures can be found in the Ktunaxa people.

## THE KTUNAXA SPEECH COMMUNITY

The Ktunaxa people have traditionally occupied plateau and prairie lands in British Columbia, Alberta, Montana, and Idaho (Schaeffer, 1940; Smith, 1986). The Ktunaxa Nation Council (KNC) defined the heartland of the Ktunaxa people in reference to geographical features rather than Federal, Provincial, or State demarcations; 70,000 sq. kilometers adjacent to the Kootenay and Columbia rivers, and the Arrow Lakes and Flathead Lake (www.ktunaxa.org). The 2001 Canadian Census identified Ktunaxa 220 speakers, while the 2006 census identified 200 speakers (Statistics Canada; 2001, 2006). The KNC Traditional Knowledge and Language program presented much lower numbers, however, asserting 24 fluent speakers – all of whom are 65 years or older (Quinn, 2010).

The Ktunaxa language speaking collective can be defined as a speech community. Morgan (2004) referred to these communities as distinctive collective consciousness capable of unifying under conditions of crisis. She added speech communities are central in meaning making studies because they provide a prolonged set of social interactions from which to evaluate shared beliefs and value systems. When perturbed by some externality, these communities provide a useful means of examining "identity, ideology, and agency" (p. 3). For Ktunaxa, externalities include a wide-ranging continuum of impacts accompanying the arrival of English. Some impacts seated English equally with Ktunaxa, while others were designed to replace it completely. Ktunaxa is a linguistic isolate; it is not related to the languages of First Nations peoples that surround the Ktunaxa (Bouchard & Kennedy, 2005). This isolation further disadvantages the Ktunaxa in maintaining a continuously fluent corpus of speakers.

A KNC (2011) report drew from extensive qualitative interviews and the Ktunaxa Nation Census (2010). It identified several specific obstacles to Ktunaxa language survival. One concern mirrored UNESCO endangerment; there is decreasing cultural interaction between younger beginner speakers, intermediate, and elderly fluent speakers. As a result, recent language efforts focused on an increased cross-generational interaction using a differentiated collection of digital resources. These high quality multimedia resources sought to reestablish intergenerational bonds between Ktunaxa children, parents, and grandparents. As one Ktunaxa language teacher mentioned, "Our kids have PlayStations and Xboxes, they have DVDs and Satellite Dishes. They have phones that get the internet. Our schools have cassette tapes and coloring books. Which one would they rather spend their time on?" (J. Louie, personal conversation, March 2010) With that in mind efforts are being made to bring speakers together in the digital realm. Thus, sophisticated websites with Flash content and PDF versions of language lessons and dictionaries are posted at [ktunaxa.org](ktunaxa.org) and [wupnik-natanik.com](wupnik-natanik.com). Wupnik' Natanik translates to "New Times", and is often equated with technology. The website was designed to offer an alternative online social network for Ktunaxa language speakers. Although a similar group existed on Facebook, Wupnik' Natanik offered a graphical user interface unincorporated historic photos and cultural themes, as well as custom designed fonts for Web publishing. Once members create profiles they are able to post status updates, send messages to other members, and access various Ktunaxa language learning resources.

**COLLECTIVE INTELLIGENCE**

So what qualifies this exertion as Collective Intelligence (CI) research? The answer can be found in the motivations and actions of the website members. While a significant amount of Ktunaxa Nation Council publicity was directed towards reviving the Ktunaxa language, little progress was made using conventional heritage language or second language resources available to community members. Local language classes tended to be underutilized and print dictionaries were organized by Ktunaxa phrases, making it cumbersome to locate English words a beginner speaker may want to translate. Most substantive language learning occurs in the primary schools with student ages ranging between four and ten years of age. When adults did partake in language classes their "lousy language learning abilities" (Trudgill, 2001) resulted in a sort of linguistic simplification. Lupyan and Dale (2010) asserted morphologically complex languages with histories of adult learning tend to become "morphologically simpler, less redundant, and more regular/transparent" (p. 2). Fluent speakers grew increasingly frustrated with poor adult beginner pronunciation of diphthongs and fricatives and their inability to accurately differentiate between regular and glottal consonants. These factors combined with small Indian Reserve populations and an English-only administration made it increasingly difficult for the Ktunaxa speech community to preserve its specificity. These tensions resulted in increased disorder, not only in communication, but in the cultural nexus as well. This interconnection binds the Ktunaxa with their identity, and increases in cultural entropy represent a significant "common problem." Adopting a social networking approach has provided contemporary speakers low cost and easily transferrable structuration resources.

CI has been defined by Brown and Lauder (2000) as "empowerment through the development and pooling of intelligence to attain common goals or resolve common problems" (p. 234). In complexity terms, Lacey (1988) viewed the collective's ability to face shared problems and to develop solutions as collective negotiations between complex social systems and the adaptations required to maintain stability far-from-equilibrium. If we define the common problem facing the Ktunaxa speech community as language death, accompanied by significant damage to self-identity, then collective actions attempting to reverse language death could be cast as CI. Other authors have presented adaptations of CI in online and trauma settings, where Morgan's (2004) unifying and collective agency resources are highly accessible and decentralized. Hiltz and Turoff (1993) defined collective intelligence in terms of online distributed problem solving, while Vieweg, Palen, Liu, Hughes, and Sutton (2008) applied online collective problem solving to disaster or trauma scenarios. Specifically, they found crises prompted a subset of agents to disseminate information in an emergent and collective manner, rather than an orchestrated or centralized one. This method of empowering individual problem solvers to access the collective problem solving resources would seem appropriate given the historical trauma the Ktunaxa speech community has survived. Brown and Lauder (2000) highlighted empowerment as "exercised through the development of the art of conversation which gives an authentic voice to all constituencies in society" (p. 238). Similarly, Avis (2002) presented a variation of CI emphasizing the removal of demarcating restrictions on communicative actions. His argument reasoned the maximization of human potential, and therefore intelligent collective action would transcend class antagonisms and achieve "collective intelligence through open dialogue" (p. 320). The relatively low transaction costs and public access to the Wupnik' Natanik facilitates this kind of unrestricted access.

Malone, Laubacher, and Dellarocas (2009) established a useful atomism for investigating CI. They identified the genetics of seemingly intelligent collaboration as stemming from *Who*, *Why*, *What*, and *How* factors that support this type of study. In this study the *Who* element is the Ktunaxa online speech community. This is a partial network of the full Ktunaxa speech community, and overlaps with various discursive structures within the Ktunaxa population. The *Why* element is to combat the increasing cultural entropy associated with the endangerment of Ktunaxa as a living language. Secondary *Why* elements include the micro-motives of the individuals reconnecting with their cultural identity. The *What* and *How* elements are the communicative interactions by the individuals, facilitated by the website, and manifested by the collective online community. Participation in the Wupnik' Natanik network demonstrated the Malone *et al's* (2009) CI genetics. The communicative actions of the individual participants constituted the building blocks of collective community agency. This agency, whose aim is to reduce cultural entropy, is voluntary and presented as a way to add to the cultural identity of the Ktunaxa Nation. Members were regularly encouraged to make suggestions as to improved reliability of cross-platform/cross–browser resources as well as the language or cultural content. Generally, there were no authorities identified by the network. The interactional media included privately posted messages between members, member status updates, member blogs, and messages to member lists, as well as participation in language lessons and collaborative authoring found in blogs or photo tagging. Another decentralizing characteristic was the pledge that any member could post audio or video content, or provide links to external language resources. The only requirement potential members were asked to abide by was a constructive and proactive contribution to the Ktunaxa language.

This atmosphere, when combined with members' dedication to the language, provided ample grounds for Bonabeau's (2000) notion of cultivating CI as an alternative to individual decision-makers acting on biased assumptions, even if those assumptions appear rational and optimal to the individual decision making agents. Bonabeau noted, "Harnessing the collective intelligence of those who have the necessary information for the benefit of those who must take action in the field can be a surer path to success than the use of top–down, template–based decisions" (p. 47). This proposed network with a loose hierarchical structure was intended to aid the decision–making agents by acting as an information broker. The intent was to allow interests to connect with information in a culturally relational environment. The environment did not disappoint. Fifteen original invitations were sent out by e-mail, which grew rapidly to eighty members. The sample included relatively equal representations of age groups from teenagers to elders. Additionally, a number of geographically isolated individuals were able to establish communication lines with resource persons otherwise unavailable to them. From their conversations several sets of patterned behaviors established emergent self-organized behavior.

## **RE-DIFFERENTIATION**

The Ktunaxa people have endured a protracted struggle to maintain stewardship of their resources, speech and otherwise. Like many other Canadian First Nations people, they were subject to educational assimilation at the hands of missionary educators (Mugocsi, 1999). The Federal Government of Canada constructed St. Eugene's Mission School in 1910 (Ktunaxa Nation, 2007). The federal government funded its administration, first by the Oblates of the Mary Immaculate, then by the Anglican Church of Canada. St. Eugene's was typical of the Canadian Indian Residential School System experience in one important respect: the faculty, staff, and clergy actively and aggressively dispossessed the First Nations students of their language (Aboriginal Healing Foundation, 2008). In addition to Ktunaxa students, the faculty and staff attempted to reprogram Okanagan, Shuswap, and Blackfoot children as farmers and animal husbandry specialists (Ktunaxa Nation, 2007; Quinn, 2010), effectively removing the specialized occupational skills and Indigenous epistemologies. The faculty and staff at the St Eugene's Mission school convinced students to abandon Ktunaxa as a primary means of family discourse (M. Tenesse, personal conversation, March 2010). These actions had a profound impact on the family structure of the Ktunaxa, severing entire generations of children from their parents and grandparents. Yet the youngest of the speakers still carry words home and attempt to use them.

Indigenous scholars Duran (2006) and Yellow Horse Brave Heart (2003) argued this collective experience is a "Soul Wound" or "Unresolved Historical Grief", respectively. Generally these theories provide an account of resilience from collective trauma. Eyerman (2001) argued this trauma is a cultural process "mediated through various forms of representation and linked to the reformation of collective identity and the reworking of collective memory" (p. 1). He referenced Alexander's (2004) notion of the collective trauma process, where post-perturbation agent actions reveal a crisis of internal meaning and group self-identity. He described resilience as an articulating discourse of competing collective identity mediation and a thoughtful development of alternative reactive voices. It was, he noted, "a process that aims to reconstitute or reconfigure a collective identity through collective representation, as a way of repairing the tear in the social fabric… reinterpreting the past as a means toward reconciling present/future needs" (p. 4). Cultural trauma theorists intimately link resilience with collective memories held by individual community members, which are transferred between generations via commemorations "establish a shared past, or through discourses more specific to a particular group or collective. This socially constructed, historically rooted collective memory functions to create social solidarity in the present" (p. 8).

Horsethief (2011) presented a model of post-trauma language revival based on Lipman-Blumen's (1973) crisis-initiated dedifferentiation. Lipman-Blumen's theory of role change provided a framework for modeling community response to perceived crises. Turner (1990) further added crisis conditions could precipitate a dedifferentiation, "as roles assimilate elements from other roles, followed by reconfiguration in a different pattern as the crisis subsides" (p. 98). According to Lipman-Blumen, under extreme stress a system "is more permeable to change than in periods of stability… One response of the system to such stress involves permitting and promoting maximum rational utilization of its resources, including those resources ordinarily held in reserve or even deliberately repressed" (p. 105). For the Ktunaxa dedifferentiation strategies included speaking the language out of the public ear, avoiding or evading Government Indian Agents, and practicing ceremonial or spiritual activities underground (KNC, 2009; Louie, 2010). These strategies appear to be consistent with Morgan's (2004) idea that speech communities were collectively conscious and capable of unifying under conditions of crisis (p. 3). What has happened recently, however, is a redifferentiating. This has been evidenced by the increased numbers of beginner Ktunaxa speakers, increased numbers of Ktunaxa seeking and receiving traditional names, and more conversational Ktunaxa being used in the community. This has been achieved largely by pursuing an overlap between Ktunaxa social networks and Ktunaxa cultural semantic networks. Gruber (2007) contended social networks are ecosystems of participation, "where value is created by the aggregation of many individual user contributions" (p. 1). Semantic networks, Gruber argued, are ecosystems of data, "where value is created by the integration of structured data from many sources" (p. 1). Gruber's thesis was the overlap between these domains created an optimal synergy, rich in "human participation and powered by well-structured information" (p. 1).

## **NETWORK INTELLIGENCE**

The Ktunaxa have recently witnessed a relative reduction in exogenous political obstacles preventing them from embracing their own language. As a result the speech community has started to "redifferentiate" itself. This process is simply a recovering of the roles changed under crisis-motivated dedifferentiation. Generally the language is again spoken in public, there are no more "Indian Agents", and some ceremonies have been opened to the public eye. In other cultural settings younger community members are promoted through the ranks to fill the vacant seats at ceremonies and gatherings. In the specific context of social networks there are distinct collaborative strategies emerging from the shadows of the Residential Schools. If we view the Ktunaxa speech community as a social network, the collective response to the perturbation of its stable language use equilibrium exhibits Christakis and Fowler's (2009) notion of network intelligence. This form of collaborative intelligence adds to or complements individual intelligence, enables network components to blueprint themselves, and empowers social networks to self-replicate in such a way that generated knowledge is passed on to future iterations (pp. 290-291). The analytical usefulness of this complex social network behavior is twofold: it allows for the examination of individual motivations and communicative actions, while identifying resulting macro trends that may not be evident at the micro scale. This collective activity enables the Ktunaxa speech community to survive environmental perturbations to pass its knowledge on to the next iteration of the community. This property is a crucial component of a robust Ktunaxa social network, and represents a functioning CI negotiating a communal solution to the collective problem of language endangerment. We can envision an operationalization of collective memory as it continues to incorporate past experiences into present speech actions. It is in this way that a speech community resists entropic disorder, reviving collective artifacts to shape future conceptualizations of shared meaning, shared belief, or shared reality. Specific attempts to collectively solve the entropy problem can be identified in the Wupnik' Natanik social networking activities.

## EMERGENT THEMES

The general network conversations resulted in a number of emergent themes, including recombination of existing Ktunaxa font resources, re-establishing cultural artifacts through collaborative photo-tagging, preserving Ktunaxa's morphological complexity by emphasizing linguistic conventions and infusing context into discussions, and addressing cultural entropy resulting from historical trauma to the individual members of the speech community. Those examples began with individuals, and resulted in-group discussions. These can be interpreted as coordinated micro–motives of the speakers, and the macro–behaviors of the online speech community. In engaging in these activities the community is using the collective discursive structures to intelligently address several key issues related to endangerment. The acts present evidence of collective network intelligence as internal mechanisms within the speech community seek to ensure the cultural network will survive far-from-equilibrium. Reinforcement of these social structures represents active resistance to UNESCO endangered speech characteristics. These activities are not reducible to the actions of the individual speech community members, nor are they simple top-down coordinating rules. They arise from the self-organized emergent "norms" that individual community members recognize as critical to the survival of the language community. Following are brief descriptions of these collectively intelligent actions:

*The font*. The development of a linguistic font that can accurately represent the non-English Ktunaxa characters has vexed Ktunaxa speech community members for nearly two decades. Several fonts have been created, distributed and mapped to various keyboard layouts with no clear winner. These initiatives have been hampered by both browser/application issues and cross-platform reliability issues. The font distributed on the Wupnik' Natanik website was "Akuqlil" (Ktunaxa for writing). It was an Open Type Font that should have been easily applied across platforms and applications. This, however, was not the case. Subsequently, it was not adopted by a majority of users. Site members cited widespread reasons, including cross-platform performance issues and installation problems. Some of this technical feedback was from tech-savvy members, many of whom used the font successfully in other applications, but could not get it to display properly. What emerged was a mix of QWERTY characters appended with apostrophes for glottal sounds (i.e.: k', q', m') and three specialized Unicode characters for full glottal stops, the barred l, and drawn out vowels (ʔ, ł, and ·). The reason for the novel recombination of resources was economization of typing effort. Only the Unicode characters required a keystroke combination, and all other letters could be typed relatively easily. This prevented users from having to memorize 4-digit key alt codes (as required for the early fonts) or keyboard remapping (as used by later fonts). By using the pidgin font, an easy alternative could reliably be used between members across platforms and browsers.

*Photo-tagging*. Still images were posted in common areas for members to view, both by the site administrator and by individual members. Many of these images were historic black-and-white photos from the Ktunaxa Nation Council archives portraying a number of Ktunaxa ancestors involved in a wide range of settings and activities. Almost immediately site members began commenting on photographs. In addition to the dates and locations of photos, the names of individuals were also posted. These names were not simply the subject's English name, but also the Ktunaxa version of their English name, and their Ktunaxa name. Other specific items were also identified by name, including geographic features and personal items appearing in photographs. One member posted a photo of family members with the following message: "Does anybody know where this picture was taken?" Within a week several individuals had commented on the location – but more importantly they also commented on the name of the man in the picture. What was interesting was that his English name was provided, the Ktunaxa version of that English name, and his Ktunaxa name, as well as similar information about his wife and descendants who currently used the names. This particular photo-tagging session continued with a discussion of the proper spelling and pronunciation of both the photo subject's and his wife's Ktunaxa name. Three individuals continued the conversation about the origination of the name, and slight variations in the pronunciation depending on the particular dialect being spoken. This conversation then evolved into one of several discussions of morphology.

*Morphological structuration*. Other collaborative activities that served to reinforce the inflectional morphology of Ktunaxa included jointly authored comments on videos and language lessons. Ktunaxa, like many Native North American languages is morphologically complex, relying heavily on emic context and nuance for descriptive accuracy. This is often contrasted to weakly inflected languages like English of German, where detail is conveyed through "non-morphological devices such as word order and lexical constructions" (Lupyan & Dale, 2010, p. 2). In the context of Ktunaxa, adult learners have increasingly relied on English non-morphological devices to communicate and are finding it increasingly difficult to appreciate the importance of Ktunaxa morphological context. Several discussions about these issues were posted on the comment fields of videos shown decomposing a Ktunaxa words into rudimentary elements. Several members commented on videos, asserting that a better understanding of root words was useful. Other members with linguistics backgrounds then extended the conversation to roots, stems, and affixes necessary for complex verb construction. Several of the conversation participants then began to discuss the possibility of developing a set of grammar rules based on morphosyntax. One of the reasons given for an increased interest in grammar rules is that even beginner speakers were aware that the language was being "dumbed-down" as more "baby talkers" struggled with complex rules of inflection inherent in Ktunaxa. Without explicit mention of the UNESCO endangerment characteristics, site members were collectively solving common problems of increased disorder resulting from gaps in linguistic structure.

*Sociability.* Finally, a number of site participants expressed gratitude for having an area to discuss language and cultural issues. Many of the participants were physically disenfranchised from the community, often separated from the traditional territory by significant distances. The site gave them an easily accessible forum to participate in important discussions they would otherwise be isolated from: discussions of topics included traditional Indian Names and their histories, traditional songs and practices, and common kinship terms for relatives. The site blogs, in particular, provided the greatest amount of collaborative authoring in terms of word count and number of individuals contributing. The information being incorporated into the blogs were directly related to the issues of identity and cultural trauma. There were not necessarily new words learned, but the multitude of contributors discussed the obstacles to learning new words, supporting Eyerman's (2001) thesis that post perturbation collective resilience included reformation of the collective identity and reworking of collective memory. It also supports Alexander's (2004) thesis that individual contributions to articulating discourses contribute to collective identity mediation – which I see as central to the group problem solving of many contemporary social issues associated with collective cultural trauma. Weiss (2005) supported this avenue, when he noted blogs don't create new kinds of content, rather they automate pre-existing context. Weiss posited, "Because blogs can be interactive exchanges between writer and reader, virtually anyone comfortable with surfing the Web could now create their own online community. Blogs shrunk the gap between consumer and producer" (p. 20). The mechanics of sharing electronic ideas and opinions is not new, but delivering the means to non-technical participators "introduced that highly combustible fuel—critical mass" (p. 20). Need for belonging in a culturally supportive atmosphere is the individual Ktunaxa's way into Weiss's notion of critical mass.

## FIGURES

Figure 1 depicts the most recent Wupnik' Natanik conversations based on messages sent between members and other members, or members posting massages about media (photos, videos, or Flash modules). In the first graphic grey nodes represent site members, clear nodes represent media elements, and the edges show the relational bonds encapsulated in conversation threads. Areas of dense edges converging on a single node should be interpreted as highly connected or popular node. When the word counts of these interactions are high, then the node is acting as a hub of cultural identity interchange. Not incidentally, these are commonly interactions between linguists, cultural consultants, or elders. In this graph blogs are green nodes, media elements are grey nodes, and members are blue nodes.

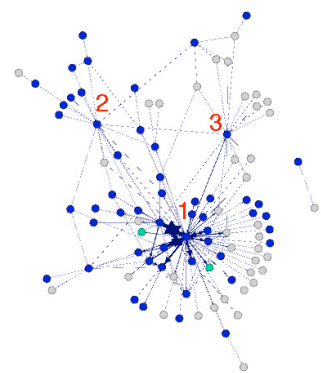

*Figure 1:   Network activity rendered by Gephi.8.*

Note nodes 1 and 2 are highly active with other nodes, and node 3 is active in photo-tagging. Also note the greatest weighted relationships are between node 1 and the several other nodes and the blogs. When the data is arranged in a circular arrangement, with the media element nodes on the left and the member nodes on the right, the effect of interacting conversation participants results in slightly more dense and clustered arrangement in the upper right quadrant. This represents a confluence of communicative actions between members and media elements. Similarly, the nodes surrounding the blogs are highly active in inter-member conversations, and blog participation. In this weighting node 2 appears less central in conversations between types of nodes (i.e.: blogs and pictures). Also of note, the relations that are indicated with arrowheads represent the discussions heavy in linguistic content (i.e.: morphology and grammar conversations). Also, the media node at the 4 O'clock is one of the pictures with significant discussion of traditional names, with emphasis on spellings across dialects.

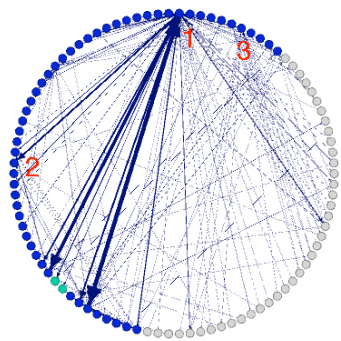

*Figure 2: Circular illustration via Gephi.8.*

**FUTURE RESEARCH POSSIBILITIES**

This paper represented the first attempt to arrange data from the Wupnik' Natanik site in a Social Network Analytic framework. The next steps will focus on specific aspects of collective network intelligence addressing the common problem of Ktunaxa language endangerment. These include identification of network leaders (indicated by high connectedness), small world architectures (represented by low geodesic distances to other nodes), and evidence of members using Schelling points (Schelling, 2006) to coordinate micro-motives for global structuration. This research also offers opportunity to investigate instances of members negotiating cultural identities online, maximizing access to other members. This idea was presented by Shirky's (2010) assertion a collaborative social network has to be engaged in more than the knowledge of the whole group to work together. For Shirky the individuals must be knowledgeable in each other's contributions to the collective problem. Similarly, there appears to be opportunity to identify Surowiecki's (2004) "wisdom of crowds effect" as non-technical community members recombine technical resources in novel ways. This can be pursued in terms of fonts, cultural information, or emergent grammar rules. Such activities may generate successful working models that have thus far eluded font designers, cultural consultants, or curriculum designers.

**CONCLUSIONS**

The data presented here provides evidence of complex emergence in a social network framework. The events of the past century have dispossessed many Native American and First Nations peoples of their language resources. The Ktunaxa speech community has developed novel approaches resisting simplification and extinction of their language, and concurrently their culture. Applying Ashby's (1956) law of requisite variety to Language as a Complex Adaptive System requires the speech community to be as complex as the environment it resides in. This is the uncoordinated and decentralized empowerment central to key to robustness in the Ktunaxa cultural identity. Unfortunately for the Ktunaxa their mother tongue has been continually simplified by the incursion of English into the every day usage community life. This trend, coupled with a staggering rate of mortality of the most fluent in Ktunaxa has qualified the language as endangered by UNESCO definitions. The online community Wupnik' Natanik has collectively resuscitated and reversed this trend, engaging in activities designed to offset the UNESCO traits: an increase in daily usage, an increase in generational transfer, and an increase in the complexity of registers. Self-organized regular communications between speakers of an array of ages is evident. As is the dedication to the complex aspects of Ktunaxa, including complex morphological structures and verb constriction grammars. The online Ktunaxa speech community's collective authoring and substantive communications regarding morphological structures evidence these autopoietic activities. Both activities represent individuals pursuing functional variations in their communicative frameworks to empowerment future speakers to resolve the common problems of language endangerment. These uncoordinated collective speech acts have found a way to "breathe" together. This collectively intelligent breathing of life back into the language, online or otherwise, allows Ktunaxa speakers to breath life back into their cultural identity.